\documentclass[aps,prc,twocolumn,showpacs,amsmath,amssymb,nofootinbib,superscriptaddress]{revtex4-1}
\usepackage{graphicx} 
\usepackage{dcolumn}

\begin{document}
\title{High-spin structures of $^{88}_{36}$Kr$_{52}$ and $^{89}_{37}$Rb$_{52}$: 
Evolution from collective to single-particle behaviors 
}
\author{A.~Astier}
\author{M.-G. Porquet}
\affiliation{CSNSM, IN2P3-CNRS and Universit\'e Paris-Sud, B\^at 104-108,
F-91405 Orsay, France}
\author{Ts.~Venkova}
\affiliation{CSNSM, IN2P3-CNRS and Universit\'e Paris-Sud, B\^at 104-108,
F-91405 Orsay, France}
\affiliation{INRNE, BAS, 1784 Sofia, Bulgaria}
\author{G.~Duch\^ene}
\affiliation{Universit\'e de Strasbourg, IPHC, 23 rue du Loess, F-67037 Strasbourg, France}
\affiliation{CNRS, UMR7178, F-67037 Strasbourg, France}
\author{F.~Azaiez}
\altaffiliation{Present address: IPNO,  IN2P3-CNRS and Universit\'e Paris-Sud, 
F-91406 Orsay, France.}
\affiliation{Universit\'e de Strasbourg, IPHC, 23 rue du Loess, F-67037 Strasbourg, France}
\affiliation{CNRS, UMR7178, F-67037 Strasbourg, France}
\author{D.~Curien}
\affiliation{Universit\'e de Strasbourg, IPHC, 23 rue du Loess, F-67037 Strasbourg, France}
\affiliation{CNRS, UMR7178, F-67037 Strasbourg, France}
\author{I.~Deloncle}
\affiliation{CSNSM, IN2P3-CNRS and Universit\'e Paris-Sud, B\^at 104-108,
F-91405 Orsay, France}
\author{O.~Dorvaux}
\author{B.J.P.~Gall}
\affiliation{Universit\'e de Strasbourg, IPHC, 23 rue du Loess, F-67037 Strasbourg, France}
\affiliation{CNRS, UMR7178, F-67037 Strasbourg, France}
\author{N.~Redon}
\affiliation{IPNL, IN2P3-CNRS and Universit\'e Claude Bernard, F-69622 Villeurbanne Cedex, France} 
\author{M.~Rousseau}
\affiliation{Universit\'e de Strasbourg, IPHC, 23 rue du Loess, F-67037 Strasbourg, France}
\affiliation{CNRS, UMR7178, F-67037 Strasbourg, France}
\author{O.~St\'ezowski}
\affiliation{IPNL, IN2P3-CNRS and Universit\'e Claude Bernard, F-69622 Villeurbanne Cedex, France}

\date{Received: date / Revised version: date}
\date{\hfill \today}

\begin{abstract}
The high-spin states of the two neutron-rich nuclei, $^{88}$Kr$_{52}$ 
and $^{89}$Rb$_{52}$ have been studied from the $^{18}$O + $^{208}$Pb 
fusion-fission reaction. Their level schemes were built from triple
$\gamma$-ray coincidence data and $\gamma-\gamma$ angular correlations were
analyzed in order to assign spin and parity values to most of the observed states. 
The two levels schemes evolve from collective structures to single-particle
excitations as a function of the excitation energy. Comparison with results of
shell-model calculations gives the specific proton and neutron configurations
which are involved to generate the angular momentum along the yrast lines.
\end{abstract} 


\maketitle

\section{Introduction}

Many years ago, the systematics of nuclear ground state properties of long
sequences  of $_{38}$Sr, 
$_{37}$Rb and $_{36}$Kr isotopes were measured by laser
spectroscopy~\cite{bu90,th81,ke95}. They show that from $N=50$, the magic
gap, to $N=58$, the mean square charge radii grow slowly, indicating small
departures from spherical shape. This was confirmed by the behavior of the
low-energy excitations of these nuclei. For instance the first two excited 
states of the even-$Z$  
$N=52$ isotones with $Z=32-38$ are those expected for a spherical harmonic
vibrator: The 2$^+_1$ energies are less than 1~MeV and the $E(4^+_1)/E(2^+_1)$ 
ratios are close to 2. It is well known
that along the yrast lines, the purely vibrational picture exhibited by nuclei 
near closed shells is perturbed by the admixture of other degrees of 
freedom, particularly the broken-pair states which provide 
large gain of angular momentum. That gives access to many proton and neutron 
configurations involving the high-$j$ orbits close to the Fermi levels.

Nuclei with few protons above $Z=28$ and few neutrons above $N=50$ provide good
testing grounds for the spherical shell model. Particularly, their high-spin 
states which are generated by aligning the angular momenta of broken nucleon 
pairs, can be clearly used to test some of the two-body matrix elements which 
are needed to predict the behavior of $^{78}$Ni and its neighbors. 

The present work reports on the study of the high-spin states of $^{88}_{36}$Kr 
and $^{89}_{37}$Rb using the $^{18}$O + $^{208}$Pb 
fusion-fission reaction. The level scheme of $^{88}$Kr was extended up to
8-MeV excitation energy and several new transitions were added to the one of 
$^{89}$Rb. Moreover $\gamma-\gamma$ angular correlations were
analyzed in order to assign spin and parity values to most of the high-spin 
states of these two $N=52$ isotones. The comparison of the two level schemes with the
predictions of shell-model calculations shows a good agreement. The analyses of the
wave functions indicate that, except those of the first excited states which are 
very mixed, they are dominated by specific proton and neutron configurations, i.e., 
single-particle degrees of freedom of both protons and neutrons play a dominant
role.  


\section{Experimental details}
\subsection{Reaction, $\gamma$-ray detection and analysis\label{exp}}

We have used the $^{18}$O + $^{208}$Pb reaction at 85-MeV incident 
energy. The beam was provided by the Vivitron accelerator of IReS 
(Strasbourg). A 100 mg/cm$^{2}$ target of $^{208}$Pb 
was used to stop the recoiling nuclei. The $\gamma$ rays were detected 
with the Euroball array~\cite{si97}. The spectrometer contained 15 
Cluster germanium detectors placed in the backward hemisphere with 
respect to the beam, 26 Clover germanium detectors located 
around 90$^\circ$, and 30 tapered single-crystal germanium detectors 
located at forward angles. Each Cluster detector consists of seven 
closely packed large volume Ge crystals~\cite{eb96} and each 
Clover detector consists of four smaller Ge crystals~\cite{du99}. 

The data were recorded in an event-by-event mode with the 
requirement that a minimum of three unsuppressed Ge
detectors fired in prompt coincidence. A set of $4 \times 
10^{9}$ three- and higher-fold events was available
for the subsequent analysis. The offline analysis consisted 
of both multi-gated spectra and three-dimensional 'cubes' built 
and analyzed with the Radware package~\cite{ra95}.

More than one hundred nuclei are produced at high spin in 
such fusion-fission experiments, and this gives several thousands 
of $\gamma$ transitions which have to be sorted out. Single-gated
spectra are useless in most of the cases. The selection of one 
particular nucleus needs at least two energy conditions, implying 
that at least two transitions have to be known. 
This is the case of the yrast bands of the two nuclei of interest, 
$^{88}$Kr and $^{89}$Rb.
We have nevertheless checked that their new transitions 
are detected in coincidence with those emitted by complementary 
fragments~\cite{ho91,po96}.   
For the reaction used in this work, we have studied many pairs of 
complementary fragments with  known $\gamma$-ray cascades
to establish the relationship between their number of protons 
and neutrons~\cite{po04}. The sum of the proton numbers of complementary 
fragments has been found to be always the atomic number of the 
compound nucleus, Z = 90. The total number of emitted neutrons 
(sum of the pre- and post-fission neutrons) is mainly 4, 5, and 6.

\subsection{$\gamma$-$\gamma$ angular correlations \label{correl}}
In order to determine the spin values of excited states,
the coincidence rates of two successive $\gamma$ transitions are 
analyzed as a function of $\theta$, the
average relative angle between the two fired detectors.
The Euroball spectrometer had $C^{2}_{239}$~=~28441 
combinations of 2 crystals, out of which $\sim$ 2000  
involved different values of relative angle within 2$^\circ$. 
Therefore, in order 
to keep reasonable numbers of counts, all the angles have been 
gathered around three average relative angles : 22$^\circ$, 46$^\circ$, 
and 75$^\circ$. 

The coincidence rate is increasing between 0$^\circ$ and 
90$^\circ$ for the dipole-quadrupole cascades, whereas it decreases for 
the quadrupole-quadrupole or dipole-dipole ones. More precisely, the 
angular correlation functions at the three angles of interest were 
calculated for several combinations of spin sequences, corresponding to 
typical multipole orders and their ratios are given in Table \ref{correl_th}. 
In order to check the method, angular correlations of transitions 
belonging to the yrast 
cascades of the fission fragments having well-known multipole orders 
were analyzed and the expected values were found in all cases.  
\begin{table}[!h]
\caption{Values of the ratios of the angular correlation functions, 
W($\theta$)/W(75$^\circ$), computed for several combinations of 
spin sequences and multipole orders ($Q$ = quadrupole, $D$ = Dipole).
}
\label{correl_th}
\begin{ruledtabular}
\begin{tabular}{cccc}
Spin sequence & Multipole  &W(22$^\circ$)/W(75$^\circ$)&W(46$^\circ$)/W(75$^\circ$)\\
$I_1-I_2-I_3$ & orders& &  \\
\hline
6--4--2& $Q-Q$ &  1.13& 1.06		\\
6--5--4& $D-D$ &  1.06& 1.03		\\
5--4--2& $D-Q$ &  0.92& 0.96		\\
5--4--2& $(D+Q)\footnotemark[1]-Q$ &  0.73& 0.88	\\
5--4--2& $(D+Q)\footnotemark[2]-Q$ &  1.63& 1.35	\\
4--4--2& $D-Q$ &  1.25& 1.13		\\
\end{tabular}
\end{ruledtabular}
\footnotetext[1]{with a mixing ratio $\delta = -1$.}
\footnotetext[2]{with a mixing ratio $\delta = +1$.}
\end{table}


\section{Experimental results}
\subsection{Level scheme of $^{88}_{36}$Kr\label{Kr88}}
The transitions deexciting the first two states of the yrast line of 
$^{88}$Kr were identified many years ago from the $\beta$ decay of 
$^{88}$Br~\cite{nndc}, establishing the 2$^+_1$ state at 775 keV and the 4$^+_1$
state at 1643 keV. More recently five new
transitions deexciting medium-spin states were identified, following the
spontaneous fission of $^{248}$Cm~\cite{rz00}. 

\begin{figure}[!h]
\includegraphics[width=8.6cm]{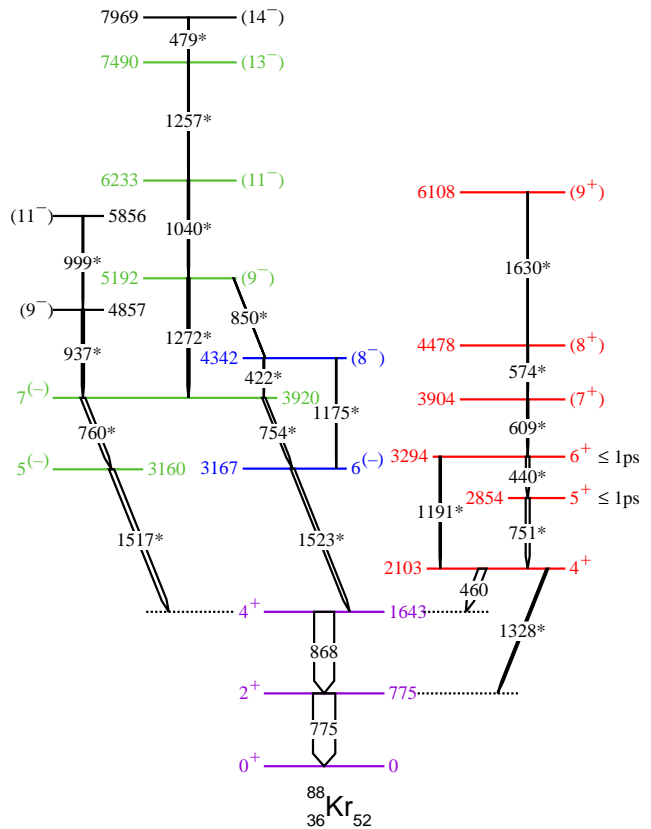}
\caption[]{(Color online) Level scheme of $^{88}$Kr established in this work. 
Transitions, which are new or are placed differently as compared to 
Ref.~\cite{rz00}, are marked with a star. The width of the arrows is 
proportional to the $\gamma$-ray intensity. The color code is the same 
as that of Fig.~\ref{88Kr_SM_exp}.
}
\label{schema88Kr}
\end{figure}

The spectrum doubly-gated by the first two transitions of 
$^{88}$Kr (at 775 and 868 keV) and built from our data set, 
contains, in addition to these five transitions, a lot of other 
$\gamma$ lines which are less intense. Therefore, we have analyzed 
all their coincidence relationships in order to build the level 
scheme which
displays two independent parts lying above the 4$^+_1$
state at 1643 keV (see Fig.~\ref{schema88Kr}).

\begin{figure}[!h]
\includegraphics*[width=8.5cm]{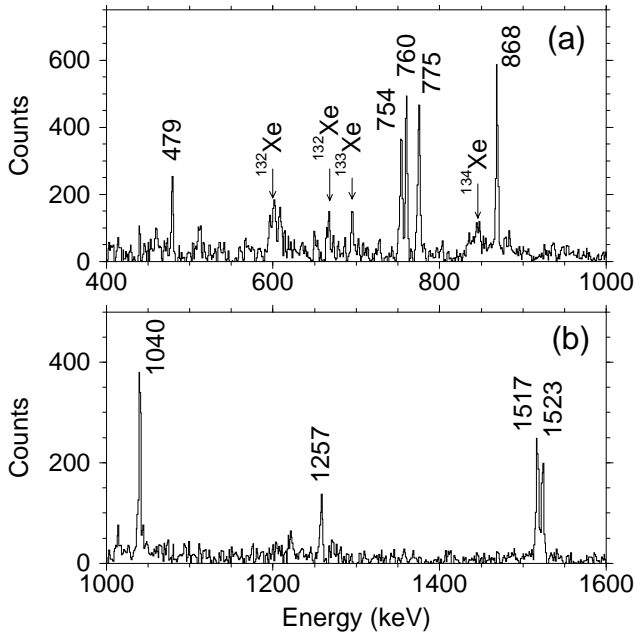}
\caption[]{Spectrum of $\gamma$ rays in coincidence with two transitions of $^{88}$Kr, the 1272-
and the 868- or 775-keV transitions,  (a) low energy part and (b) high energy part.
Transitions emitted by $^{132-134}$Xe, the complementary 
fragments of $^{88}$Kr, are labeled. 
}
\label{spectre1272}
\end{figure}

The 3920-keV level, belonging to the left part, is linked to the 4$^+_1$ state by means of two parallel
cascades, 760--1517~keV on the one hand and 754--1523~keV on the other hand. Because 
of their relative intensities, the order of the transitions
of both cascades has to be inverted as compared to the previous 
work~\cite{rz00}. Moreover the fact that the 1175-keV transition is coincident
with the 1523-keV transition and not with the 754-keV one confirms the order of
the 754--1523~keV cascade. 
Two sets of coincident transitions are located above the 
3920-keV level, (i) one set with two transitions of similar energies (937 and 999~keV) 
and (ii) one set with four transitions.
An example of doubly-gated spectra showing the transitions belonging to this
cascade is shown in Fig.~\ref{spectre1272}. 

\begin{figure}[!h]
\includegraphics*[width=7.5cm]{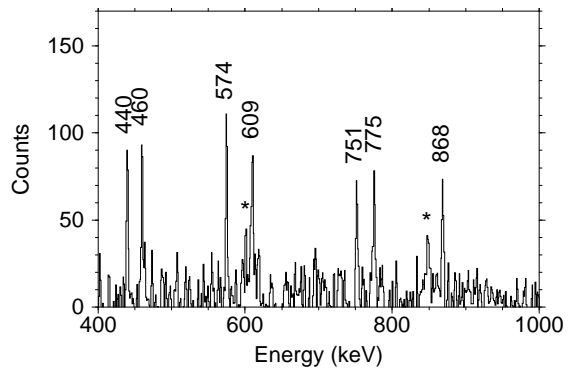}
\caption[]{Spectrum of $\gamma$ rays in coincidence with two transitions of 
$^{88}$Kr, the 1630-  and the 868- or 775-keV transitions, showing the 
transitions belonging to the right part of the level scheme. The peaks marked
with a star are contaminants.
}
\label{spectre1630}
\end{figure}

The right part of the level scheme is built on the 2103-keV level. It mainly
decays to the 4$^+_1$ state by means of the 460-keV transition. Its link
to the 2$^+_1$ state, newly observed in this work, is much weaker in intensity.
The spectrum doubly-gated by the 1630-keV transition, located at the top of this
new branch, and either the 868- or the
775-keV one is shown in Fig.~\ref{spectre1630}. 

Angular correlations of successive $\gamma$ rays have been extracted for the 
most intense transitions of $^{88}$Kr, provided that the $\gamma_i-\gamma_j$ 
coincidence of interest is not polluted. The experimental results are given in
Table~\ref{correl_Kr}.
\begin{table}[!h]
\caption{Ratios of the coincidence rates between the low-lying $\gamma$ rays of $^{88}$Kr.}
\label{correl_Kr}
\begin{ruledtabular}
\begin{tabular}{ccc}
$\gamma$--$\gamma$ coincidence & W(22$^\circ$)/W(75$^\circ$)$^{(a)}$ & W(46$^\circ$)/W(75$^\circ$)$^{(a)}$ \\
\hline
775--1523  &  1.09(9)    & 1.05(6) 	\\
868--1523  &  1.2(1)     & 1.1(1)  	\\
775--760   &  1.17(9)    & 1.03(6) 	\\
775--1517  &  0.88(8)    & 0.95(6) 	\\
1523--754  &  0.89(8)    & 0.95(6) 	\\
&&\\
775--460  &   1.23(8)	& 1.15(6) 	\\
868--460  &   1.18(8)	& 1.06(6) 	\\
775--440  &   0.82(7)	& 0.92(5) 	\\
460--440  &   0.92(7)	& 0.94(5) 	\\
751--440  &   1.09(7)	& 1.05(6) 	\\
\end{tabular}
\end{ruledtabular}
\footnotetext[1]{The number in parentheses is the error in the 
last digit.}
\end{table}

Knowing that the 775- and 868-keV transitions have a stretched quadrupole
character~\cite{rz00}, these angular-correlation results show that both 
the 1523- and 760-keV transitions are
quadrupole, while the 1517- and 754-keV ones are dipole. Thus the spin values
of the 3160-, 3167- and 3920-keV states are 5, 6, and 7, respectively. 

Regarding the right part of the level scheme, the 775--460 and 868--460 
angular-correlation
values would lead us to conclude that the 460-keV transition has a stretched 
quadrupole character, implying a $I^\pi=6^+$ assignment for the 2103-keV level.
Such a result is at variance with its decay to the 2$^+_1$ state (see
Fig.~\ref{schema88Kr}). Thus the 460-keV transition is a $\Delta I=0$ dipole
transition, since such a transition gives comparable results as a $\Delta I=2$
quadrupole one (see Table~\ref{correl_th}). It is worth noting that the 2103-keV
state has been populated with $L=3$ or 4 in a transfer reaction~\cite{nndc},
implying that it has either $I^\pi=3^-$ or $I^\pi=4^+$. The first value was
already eliminated in Ref.~\cite{rz00} because it would imply a strong $E1$
decay to the 2$^+_1$ state, at variance with the experimental results.
Thus, all these arguments result in $I^\pi=4^+$ for the 
2103-keV state. Finally, the last results of Table~\ref{correl_Kr} indicate 
that the 440- and 751-keV transitions are dipole. Given that the 3294-keV
state is directly linked to the 2103-keV state by means of the 1191-keV
transition, this defines unambiguously the spin and parity values of the 
2855- and 3295-keV states, $I^\pi=5^+$ and 6$^+$, respectively.

\begin{figure}[!h]
\includegraphics*[width=8cm]{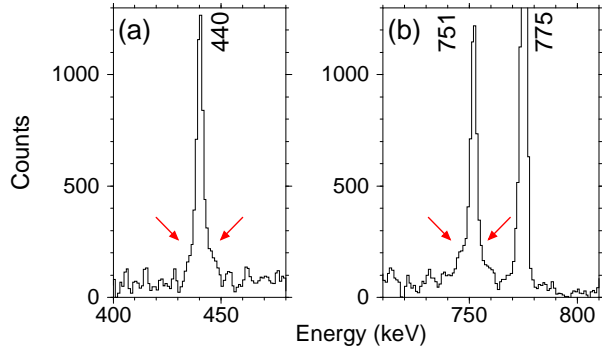}
\caption[]{(Color online) Spectrum of $\gamma$ rays in coincidence with the 460- and 
868-keV transitions. (a) the red arrows show the broadening of the 440-keV 
$\gamma$ line. (b) the red arrows show the broadening of the  751-keV 
$\gamma$ line.
}
\label{pics_larges}
\end{figure}

It is important to note that the 440- and 751-keV $\gamma$ lines exhibit 
energy broadenings in spectra gated by transitions located below 
[see the red arrows in Figs.~\ref{pics_larges}(a) and (b)]. Such shapes mean
that the 440- and 751-keV transitions are partially emitted during the 
slowing-down of the $^{88}$Kr fragments in the Pb target. Thus, the 
corresponding excited states do have half-lives $\le$~1~ps, i.e., 
$B(M1, 440~$keV$) \ge 0.45~\mu_n^2$ and $B(M1, 751~$keV$) \ge 0.09~\mu_n^2$.   

Since the first states belonging to the right part of the level scheme have a 
positive parity, we have assumed that the states of the left part do have a
negative parity (given in parentheses in Fig.~\ref{schema88Kr}). With such a
choice, the 1523-keV transition would have a $M2$ character. Several $M2$
transitions are known in the $^{85,87,91}$Rb isotopes~\cite{nndc}, they 
deexcite isomeric states with various half-lives depending on the transition
energies. Nevertheless all the $B(M2)$ values lie in a small range, 
[4.4--10]~$10^{-2}$~W.u..  Using these limits, the half-life of the 3167-keV
level would be $T_{1/2}=$~2--4~ns, which is well below the time resolution of
our apparatus~\cite{as06}.

Lastly, the spin assignments of the higher-lying states, given in 
parentheses, are based on the assumption that, in the yrast decays, spin 
values increase with excitation energy.  
We have gathered in Table~\ref{gammas_Kr88} the properties of all the
transitions assigned to $^{88}$Kr from this work.
\begin{table}[!h]
\caption{Properties of the transitions assigned to $^{88}$Kr observed in this work.}
\label{gammas_Kr88}
\begin{ruledtabular}
\begin{tabular}{rrccc}
$E_\gamma$\footnotemark[1](keV)&$I_\gamma$\footnotemark[1]$^,$\footnotemark[2]&$I_i^\pi \rightarrow I_f^\pi$&$E_i$&$E_f$\\
\hline
421.6(4)& 3.0(15)   &  (8$^-$)   $\rightarrow$  7$^{(-)}$   &4342.2  &3920.4  \\
439.8(3)& 15(4)   &    6$^+$     $\rightarrow$  5$^+$  	  &3294.2  &2854.4  \\
459.7(2)& 34(7)   &    4$^+$     $\rightarrow$  4$^+$  	  &2103.2  &1643.5  \\
478.9(4)& 1.7(8)   &   (14$^-$)   $\rightarrow$ (13$^-$)  &7969.0  &7490.1  \\
574.4(4)& 5(2)   &    (8$^+$) 	  $\rightarrow$ (7$^+$)   &4478.1  &3903.7  \\
609.5(4)& 6(2)   &    (7$^+$) 	  $\rightarrow$ 6$^+$     &3903.7  &3294.2  \\
751.2(3)& 17(4)   &   5$^+$      $\rightarrow$  4$^+$     &2854.4  &2103.2  \\
753.7(3)& 13(3)   &   7$^{(-)}$    $\rightarrow$  6$^{(-)}$   &3920.4  &3166.7  \\
759.9(3)& 20(5)   &   7$^{(-)}$  $\rightarrow$ 5$^{(-)}$     &3920.4  &3160.5  \\
775.1(2)& 100     &  2$^+$     $\rightarrow$  0$^+$  	  &775.1   &0.0  \\
850(1)& 1.5(7)   &    (9$^-$)	  $\rightarrow$ (8$^-$)   &5192.5  &4342.2 \\
868.4(2)& 90(13)   &   4$^+$     $\rightarrow$  2$^+$     &1643.5  &775.1  \\
936.6(4)& 9.0(3)   &  (9$^-$)     $\rightarrow$ 7$^{(-)}$   &4857.0  &3920.4  \\
999.2(4)& 4(2)   &    (11$^-$) 	  $\rightarrow$ (9$^-$)   &5856.2  &4857.0  \\
1040.5(5)& 5(2)   &   (11$^-$)    $\rightarrow$ (9$^-$)   &6233.0  &5192.5  \\
1175.5(7)& 2(1)   &  (8$^-$)     $\rightarrow$  6$^{(-)}$   &4342.2  &3166.7  \\
1191.2(5)& 6(2)   &   6$^+$      $\rightarrow$  4$^+$  	  &3294.2  &2103.2  \\
1257.1(7)& 2.5(12)   & (13$^-$)   $\rightarrow$ (11$^-$)  &7490.1  &6233.0  \\
1272.1(5)& 10(3)   & (9$^-$)      $\rightarrow$ 7$^{(-)}$   &5192.5  &3920.4  \\
1328.1(5)& 7(2)   &   4$^+$       $\rightarrow$ 2$^+$     &2103.2  &775.1  \\
1517.0(4)& 21(4)   &  5$^{(-)}$     $\rightarrow$ 4$^+$     &3160.5  &1643.5  \\
1523.2(4)& 17(4)   &  6$^{(-)}$     $\rightarrow$ 4$^+$     &3166.7  &1643.5  \\
1630(1)& 1.5(7)   &   (9$^+$)    $\rightarrow$ (8$^+$)	  &6108    &4478.1  \\
\end{tabular}
\end{ruledtabular}
\footnotetext[1]{The number in parentheses is the error in the last digit.}
\footnotetext[2]{The relative intensities are normalized to $I_\gamma(775) = 100$.}
\end{table}

\subsection{Level scheme of $^{89}_{37}$Rb\label{Rb89}}

Our first results on high-spin states of $^{89}$Rb were obtained when we were
searching for the high-spin structure of the odd-odd $^{88}$Rb nucleus~\cite{po99} in the data set
registered by the Eurogam array, following the $^{28}$Si + $^{176}$Yb reaction.
The $\gamma$ rays newly assigned to its neighbor, $^{89}$Rb, were detected in coincidence 
with those of $^{109,110}$Ag~\cite{po02}, 
their main complementary fragments in that fusion-fission reaction. The level
scheme of $^{89}$Rb built at that time extended up to 5326~keV and its main
feature was the observation of the numerous transitions deexciting the 1195-keV
level, already known from the $\beta$ decay of $^{89}$Kr~\cite{he73}. Some years
later, we quoted in a publication related to the revised spin value of the
ground state of its isotone, $^{87}$Br$_{52}$, that all the low-energy states of 
$^{89}$Rb can be explained as collective states built on the deformed orbitals
lying close to the $Z=37$ Fermi level for a deformation $\epsilon \sim
+0.15$~\cite{po06}.

Later, the high-spin level scheme of $^{89}$Rb was obtained by three
groups, via various reactions induced by heavy ions~\cite{bu07,to09,pa09}. 
The highest-energy state
obtained in the first work is at 4033~keV~\cite{bu07}, while the yrast line 
was extended up to 5605~keV in the second work~\cite{to09} and 7391~keV in the
third one~\cite{pa09}. It is worth mentioning that the two latter publications
are conference proceedings, displaying only the obtained level scheme. Such a
situation is summarized in a sentence  of the abstract of the last
compilation~\cite{NDS89}, {\it "For $^{89}$Rb, little information is available
for high-spin structures"}.
The $^{89}$Rb nucleus being well produced in the $^{18}$O + $^{208}$Pb 
reaction, we have resumed its study in order to get more precise information 
on its excited states. 

\begin{figure}[!h]
\includegraphics*[width=8.5cm]{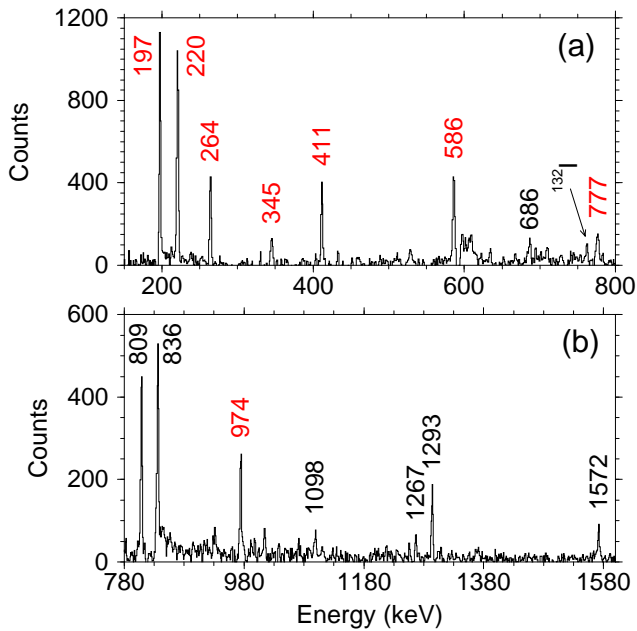}
\caption[]{(Color online) Spectrum of $\gamma$ rays in coincidence with two transitions of
$^{89}$Rb, the 1193-  and the 836- or 809-keV transitions. (a) low-energy part,
(b) high-energy part. The $\gamma$-rays involved
in the decay of the 1195-keV state are written in red. The transition 
emitted by $^{132}$I, the main complementary fragment of $^{89}$Rb in 
the $^{18}$O + $^{208}$Pb reaction, is labeled. 
}
\label{spectre1193}
\end{figure}

\begin{figure}[!h]
\includegraphics[width=8.6cm]{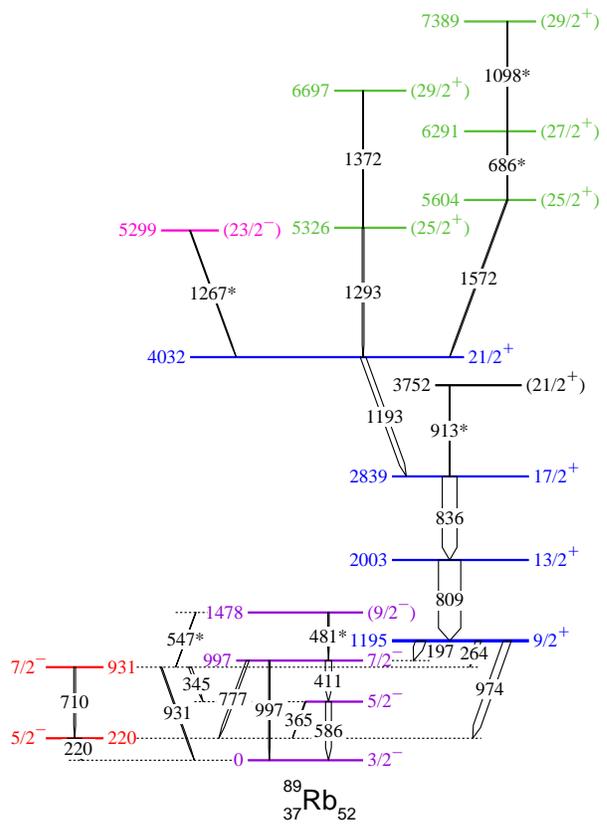}
\caption[]{(Color online) Level scheme of $^{89}$Rb established in this work. 
The width of the arrows is proportional to the $\gamma$-ray intensity.
Transitions, which are new or are placed differently as compared to 
Ref.~\cite{pa09}, are marked with a star. The 1195-keV level is isomeric, 
$T_{1/2}=15.2(28)$~ns~\cite{to09} or 8(2)~ns~\cite{pa09}. The color code is the same
as that of Fig.~\ref{89Rb_SM_exp}.
}
\label{schema89Rb}
\end{figure}

An example of doubly-gated spectrum showing most of the
transitions involved in the decay of the 1195-keV level is shown in 
Fig.~\ref{spectre1193}.
All the coincidence relationships have been carefully analyzed
in order to produce the level scheme drawn in Fig.~\ref{schema89Rb}, which is in
good agreement with the results of Ref.~\cite{pa09}, except some points which
are discussed now.
We have inverted the order of the two highest-lying transitions,
since the 686-keV transition has a higher intensity. A few other transitions
have been observed, such as the 1267- and 913-keV transitions in the high-spin
part, and the 547- and 481-keV  transitions in the low-spin part.

Angular correlations of successive $\gamma$ rays have been extracted for the 
most intense transitions of $^{89}$Rb. The experimental results are given in
Table~\ref{correl_Rb}. They indicate that these transitions can be shared out between two groups 
with different multipole orders. The 197-, 220-, 
411- and 586-keV $\gamma$ rays belonging to the first one are dipole, while 
the 809-, 836-, 974-, and 1193-keV $\gamma$ rays of the second one are
quadrupole.
\begin{table}[!h]
\caption{Ratios of the coincidence rates between the low-lying $\gamma$ rays of $^{89}$Rb.}
\label{correl_Rb}
\begin{ruledtabular}
\begin{tabular}{ccc}
$\gamma$--$\gamma$ coincidence&W(22$^\circ$)/W(75$^\circ$)$^{(a)}$&W(46$^\circ$)/W(75$^\circ$)$^{(a)}$\\
\hline
220--974  &  0.88(9)    & 0.90(6) 	\\
220--809  &  0.86(9)    & 0.95(6)	\\
220--836  &  0.89(9)    & 0.96(6) 	\\
&&\\
974--809  &  1.11(8)    & 1.05(7) 	\\
&&\\
586--809  &  0.90(8)    & 0.95(7)	\\
&&\\
411--197  &  1.08(7)    & 1.04(6)	\\
411--809  &  0.87(8)	& 0.90(8) 	\\
&&\\
197--809  &  0.88(8)	& 0.95(6)	\\
&&\\
809--836  &  1.14(9)	& 1.09(8) 	\\
809--1193 &  1.11(9)	& 1.08(8) 	\\
\end{tabular}
\end{ruledtabular}
\footnotetext[1]{The number in parentheses is the error in the 
last digit.}
\end{table}
Starting from the measured 3/2$^-$ value of the ground state~\cite{nndc}, this 
leads to the spin values of the 220-keV state (5/2), 586-keV state (5/2),  
997-keV state (7/2), 1195-keV state (9/2), 2003-keV state (13/2), 
2839-keV state (17/2), and 4032-keV state (21/2). Given that the 1195-keV state
is isomeric, $T_{1/2}=15.2(28)$~ns~\cite{to09} or 8(2)~ns~\cite{pa09}, the quadrupole 
974-keV transition has to be $M2$, implying that the 1195-keV state 
has a positive parity since the 220-keV state  
has a negative parity.In addition the spin-parity values 
of the 931-keV level, which is directly linked to the 9/2$^+$ state and to 
the 3/2$^-$ and 5/2$^-_1$ states, is unambiguously 7/2$^-$. 
All these spin and parity values are reported in
Fig.~\ref{schema89Rb} without parenthesis. 
  
We have gathered in Table~\ref{gammas_Rb89} the properties of all the
transitions assigned to $^{89}$Rb from this work. Before closing this section,
it is important to note that the $B(M2; 9/2^+ \rightarrow 5/2^-)$ value of 
$^{89}$Rb is $3.9(7)~10^{-2}~$W.u. or $7(2)~10^{-2}$~W.u. (depending on the
half-life value~\cite{to09,pa09}), i.e., in the same 
range than the ones of $^{85,87,91}$Rb (see the values given at the end of 
Sect.~\ref{Kr88}).
\begin{table}[!h]
\caption{Properties of the transitions assigned to $^{89}$Rb observed in this work.}
\label{gammas_Rb89}
\begin{ruledtabular}
\begin{tabular}{rrccc}
$E_\gamma$\footnotemark[1](keV)  &  $I_\gamma$\footnotemark[1]$^,$\footnotemark[2]  &  $I_i^\pi \rightarrow I_f^\pi$  &$E_i$&$E_f$  \\
\hline
197.4(2)& 43(9)   &     9/2$^+$	  $\rightarrow$  7/2$^-$   	&1194.6 &997.1 \\
220.5(2)& 57(8)   &     5/2$^-$   $\rightarrow$ 3/2$^-$		&220.5  &0.0    \\
263.8(2)& 23(5)   &     9/2$^+$	  $\rightarrow$  7/2$^-$   	&1194.6 & 930.7 \\
345.0(3)& 11(3)   &       7/2$^-$ $\rightarrow$  5/2$^-$ 	&930.7   &585.7  \\
365.4(4)& 3.0(15) &      5/2$^-$  $\rightarrow$  5/2$^-$    	&585.7  &220.5   \\
411.4(2)& 28(6)   &      7/2$^-$  $\rightarrow$   5/2$^-$   	&997.1  & 585.7  \\
481.4(4)& 3.0(15) &     (9/2$^-$) $\rightarrow$   7/2$^-$   	&1478.5 &997.1   \\
547.5(8)& 2(1)    &     (9/2$^-$) $\rightarrow$  (7/2$^-$)  	&1478.5 & 930.7  \\
585.7(2)& 36(7)   &      5/2$^-$  $\rightarrow$   3/2$^-$ 	&585.7  &0.0     \\
686.5(5)& 3.8(16) &    (27/2$^+$) $\rightarrow$  (25/2$^+$)	&6290.7	&5604.2  \\
710.2(4)& 8(2)    &      7/2$^-$  $\rightarrow$   5/2$^-$ 	&930.7  &220.5  \\
776.6(3)& 12(3)   &      7/2$^-$  $\rightarrow$   5/2$^-$   	&997.1  &220.5  \\
808.8(2)& 100     &     13/2$^+$  $\rightarrow$  9/2$^+$	&2003.4 &1194.6 \\
836.1(2)& 65(10)  &     17/2$^+$  $\rightarrow$  13/2$^+$ 	&2839.5 &2003.4  \\
912.8(4)& 6(2)    &    (21/2$^+$) $\rightarrow$  17/2$^+$ 	&3752.3	&2839.5 \\
930.8(4)& 6(2)    &     7/2$^-$	  $\rightarrow$   3/2$^-$ 	&930.7  &0.0    \\
974.3(3)& 34(7)   &     9/2$^+$	  $\rightarrow$  5/2$^-$  	&1194.6 &220.5 \\
997.4(4)& 6(2)    &     7/2$^-$	  $\rightarrow$  3/2$^-$  	&997.1  & 0.0   \\
1098.1(5)& 2(1)   &   (29/2$^+$)  $\rightarrow$   (27/2$^+$) 	&7388.8 &6290.7 \\
1192.6(3)& 25(5)  &     21/2$^+$  $\rightarrow$ 17/2$^+$   	&4032.1 &2839.5  \\
1266.9(4)& 3.4(17)&    (23/2)     $\rightarrow$ 21/2$^+$    	&5299.0 &4032.1 \\
1293.5(4)& 7.6(23)&   (23/2)      $\rightarrow$ 21/2$^+$    	&5325.6 &4032.1 \\
1371.6(5)& 1.8(9) &    (27/2)     $\rightarrow$ (23/2)   	&6697.2 &5325.6  \\
1572.1(4)& 4.7(19)&    (25/2$^+$) $\rightarrow$ 21/2$^+$   	&5604.2 &4032.1 \\
\end{tabular}
\end{ruledtabular}
\footnotetext[1]{The number in parentheses is the error in the last digit.}
\footnotetext[2]{The relative intensities are normalized to $I_\gamma(809) = 100$.}
\end{table}


\section{Discussion}\label{discuss}
\subsection{Generalities}\label{gene}
The characteristics of the first two excited states of many even-$Z$ $N=52$ 
isotones~\cite{nndc} are typical of a collective motion, namely that of a 
spherical harmonic vibrator. All the 2$^+_1$ excitation 
energies are less than 1~MeV, from 832~keV for $Z=38$ to 624~keV 
for $Z=32$, and all the $E(4^+_1)/E(2^+_1)$ ratios are close to 2. Nevertheless, 
such a behavior does not extend to higher angular momenta, where other sets of
states appear.

The medium-spin states of the $N=52$ isotones with $Z \sim 36$ then involve the
breaking of nucleon pairs located in the various orbits located close to the 
Fermi levels. The two valence neutrons are likely in the $\nu d_{5/2}$ and/or 
$\nu g_{7/2}$ orbits just above the $N=50$ shell gap. Regarding the
protons, three orbits located between the $Z=28$ and the $Z=50$ 
shell gaps, $\pi f_{5/2}$, $\pi p_{3/2}$, and  $\pi g_{9/2}$, are to be
considered for the pair breakings in order to increase the value of angular 
momenta. Table~\ref{break} gives various configurations of $^{88}$Kr, 
with one and two broken pairs.
\begin{table}[!h]
\caption{Various configurations, expected in $^{88}$Kr, 
with one or two broken pairs in the subshells close to the Fermi
levels, sorted by order of $I^\pi_{max}$. 
}
\label{break}
\begin{ruledtabular}
\begin{tabular}{llcc}
Neutron pair			& Proton pair 			  & $I^\pi_{max}$   &Seniority\\
				&				  &		    &(S$_\nu$+S$_\pi$)\\
\hline
&&&\\
$(\nu d_{5/2})^2$		&~~~~~---			  &  4$^+$	&    2~(2+0)	\\
~~~~~---			&$(\pi f_{5/2})^1(\pi p_{3/2})^1$ &  4$^+$	&    2~(0+2)	\\
~~~~~---			&$(\pi f_{5/2})^2$ 		  &  4$^+$	&    2~(0+2)	\\
$(\nu d_{5/2})^1(\nu g_{7/2})^1$&~~~~~---			  &  6$^+$	&    2~(2+0)	\\
~~~~~---			&$(\pi g_{9/2})^2$ 		  &  8$^+$	&    2~(0+2)	\\
&&&\\
~~~~~---			&$(\pi f_{5/2})^2(\pi p_{3/2})^2$ &  6$^+$	&    4~(0+4)	\\
$(\nu d_{5/2})^2$ 		&$(\pi f_{5/2})^1(\pi p_{3/2})^1$ &  8$^+$	&    4~(2+2)	\\
$(\nu d_{5/2})^1(\nu g_{7/2})^1$&$(\pi f_{5/2})^1(\pi p_{3/2})^1$ &  10$^+$	&    4~(2+2)	\\
$(\nu d_{5/2})^2$		&$(\pi g_{9/2})^2$ 		  &  12$^+$	&    4~(2+2)	\\
~~~~~---	&$(\pi f_{5/2})^1(\pi p_{3/2})^1(\pi g_{9/2})^2$  &  12$^+$	&    4~(0+4)	\\
$(\nu d_{5/2})^1(\nu g_{7/2})^1$&$(\pi g_{9/2})^2$ 		  &  14$^+$	&    4~(2+2)	\\
&&&\\
&&&\\
~~~~~---			&$(\pi p_{3/2})^1(\pi g_{9/2})^1$ &  6$^-$	&    2~(2+0)	\\
~~~~~---			&$(\pi f_{5/2})^1(\pi g_{9/2})^1$ &  7$^-$	&    2~(2+0)	\\
&&&\\
$(\nu d_{5/2})^2$		&$(\pi p_{3/2})^1(\pi g_{9/2})^1$ &  10$^-$	&    4~(2+2)	\\
~~~~~---	&$(\pi f_{5/2})^2(\pi p_{3/2})^1(\pi g_{9/2})^1$  &  10$^-$	&    4~(0+4)	\\	
$(\nu d_{5/2})^2$		&$(\pi f_{5/2})^1(\pi g_{9/2})^1$ &  11$^-$	&    4~(2+2)	\\
$(\nu d_{5/2})^1(\nu g_{7/2})^1$&$(\pi p_{3/2})^1(\pi g_{9/2})^1$ &  12$^-$	&    4~(2+2)	\\
$(\nu d_{5/2})^1(\nu g_{7/2})^1$&$(\pi f_{5/2})^1(\pi g_{9/2})^1$ &  13$^-$	&    4~(2+2)	\\
\end{tabular}
\end{ruledtabular}
\end{table}
Obviously some of the pure proton configurations have been already identified in 
$^{86}$Kr$_{50}$, such as the 4$^+$ state at 2250~keV, the 6$^-$ state at 4430~keV, 
or the 7$^-$ state at 4693~keV~\cite{pr04}. On the other hand, other ones, being not 
yrast, cannot be easily observed experimentally. It is likely 
the case of the 8$^+$ fully aligned state of the $(\pi g_{9/2})^2$ configuration, 
because of the 
high energy of the $\pi g_{9/2}$ orbit above the Fermi level (the
excitation energy of the 9/2$^+$ state of $^{85}_{35}$Br is 1859~keV~\cite{as06}). 
Similarly, the configuration  involving the breaking of a $(\pi f_{5/2})^2$ pair is 
lying above the yrast line, since a proton pair has to be promoted from the fully 
occupied $\pi f_{5/2}$ orbit\footnote{The proton configuration of the 
$^{88}_{36}$Kr ground state is assumed to be $(\pi f_{5/2})^6(\pi p_{3/2})^2$.}  
into the $\pi p_{3/2}$ or $\pi p_{1/2}$ subshells. Nevertheless such proton-pair
breakings must be involved in yrast states with higher senority since this is the
only way to gain extra values of angular momentum.

By using Table~\ref{break} as a guide, the main configuration of some states of 
$^{88}$Kr could be assigned. The 4$^+$ state lying at 2103~keV (see 
Fig.~\ref{schema88Kr}) likely comes from the breaking of a proton pair, 
$(\pi f_{5/2})^1(\pi p_{3/2})^1$, and the (8$^+$) state at 4478-keV excitation 
energy from the breaking of two pairs, 
$(\nu d_{5/2})^2 (\pi f_{5/2})^1(\pi p_{3/2})^1$. The 6$^{(-)}$ and 7$^{(-)}$ 
states (lying at 3167 and 3920~keV, respectively) would involve 
a different proton pair, $(\pi p_{3/2})^1(\pi g_{9/2})^1$ and 
$(\pi f_{5/2})^1(\pi g_{9/2})^1$, respectively. Adding the breaking of the neutron 
pair, either $(\nu d_{5/2})^2$ or $(\nu d_{5/2})^1(\nu g_{7/2})^1$, we could explain 
the set of states lying above 3920~keV, up to $I^\pi=13^-$.

A similar approach can be held for the high-spin states of $^{89}$Rb, a proton being added in one 
of the orbits located between the $Z=28$ and the $Z=50$ shell gaps. The states, with a senority of 
one, would be the 3/2$^-$ ground state, the 5/2$^-$ state at 220~keV and the 9/2$^+$ state at 
1195~keV, with the $(\pi p_{3/2})^1$, $(\pi f_{5/2})^1$, and $(\pi g_{9/2})^1$ configuration, 
respectively. The increase of angular momentum in each structure could be due to the breaking of 
one neutron pair or one proton pair.
For instance, the 21/2$^+$ state at 4032~keV (see Fig.~\ref{schema89Rb}) is likely the fully
aligned state of the 
$(\nu d_{5/2})^1(\nu g_{7/2})^1(\pi g_{9/2})^1$ configuration. A proton pair has then to be broken
in order to gain more angular momentum, implying that the states measured above 5~MeV excitation 
energy have likely a senority of five.   

Nevertheless, in order to have a better insight of the excitations involved in the high-spin 
states identified in $^{88}$Kr and $^{89}$Rb, particularly the respective roles
of the neutrons and the protons, the shell-model approach has to be used because of the large number of possible configurations involved in the same range 
of excitation energy.

\subsection{Results of shell-model calculations}\label{resultsSM}
We have performed shell-model (SM) calculations, using the interaction "glekpn" (described in
Ref.~\cite{ma90}) and the NuShellX@MSU code~\cite{br07}. The "glekpn" model space includes five
proton orbits ($\pi f_{7/2}$, $\pi f_{5/2}$, $\pi p_{3/2}$, $\pi p_{1/2}$, $\pi g_{9/2}$) and 
five neutron orbits ($\nu g_{9/2}$, $\nu g_{7/2}$, $\nu d_{5/2}$, $\nu d_{3/2}$, $\nu s_{1/2}$), 
which is suitable for the description of nuclei with $20~<~Z~<~50$ and $40~<~N~<~70$. 
Due to the too large dimension of the valence space when $Z~\sim~36$ and $N=52$, the $\pi f_{7/2}$ 
and the $\nu g_{9/2}$ subshells were constrained to be fully
occupied, meaning that excitations across the $Z=28$ and the $N=50$ gaps were forbidden for the
calculations we have performed in the present work. The effects
of such truncations will be appraised in the next sections, when dicussing the main results of these
SM calculations. 
\subsubsection{$^{88}$Kr}\label{SM88}

\begin{figure*}[!ht]
\includegraphics[angle=-90,width=14cm]{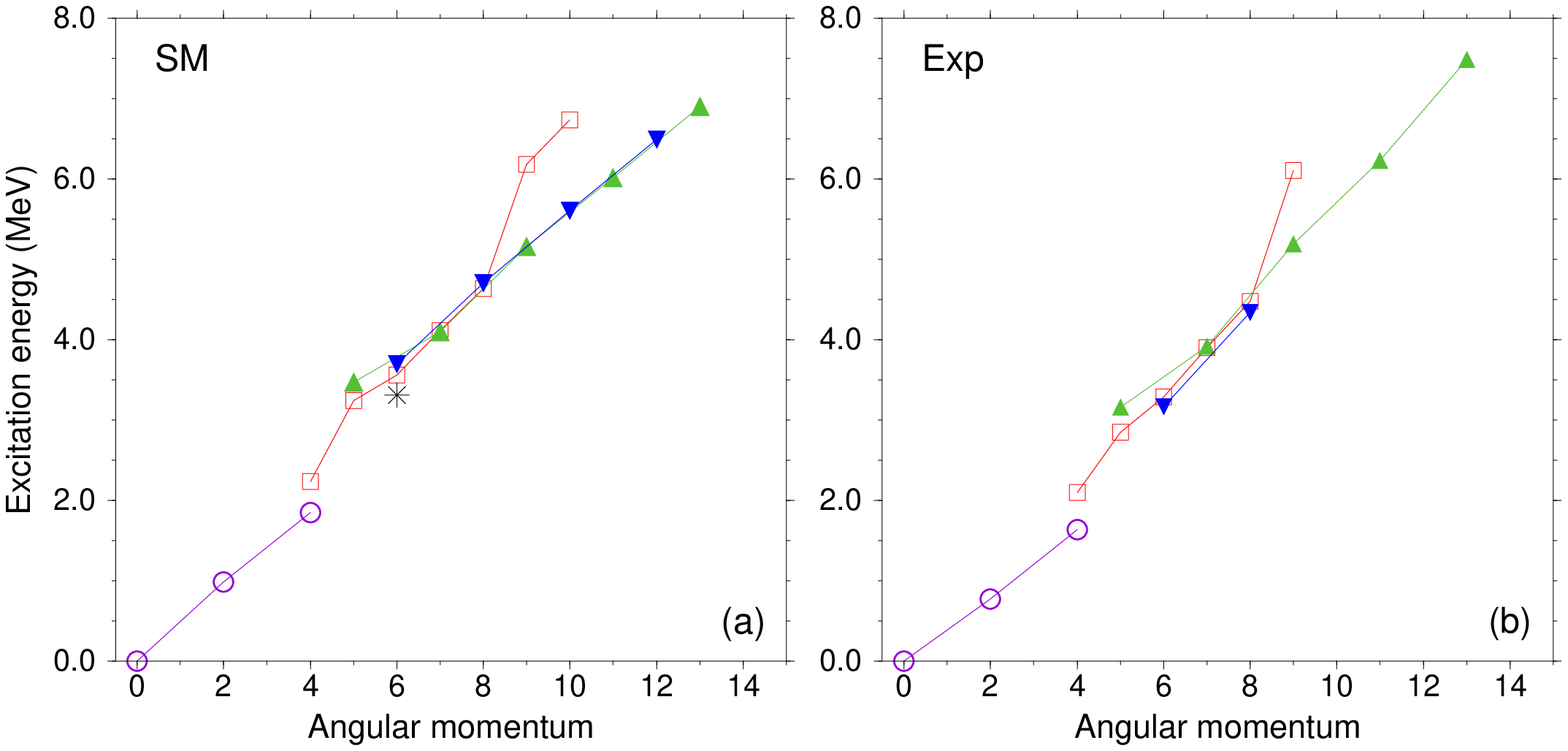}
\includegraphics[angle=-90,width=8.8cm]{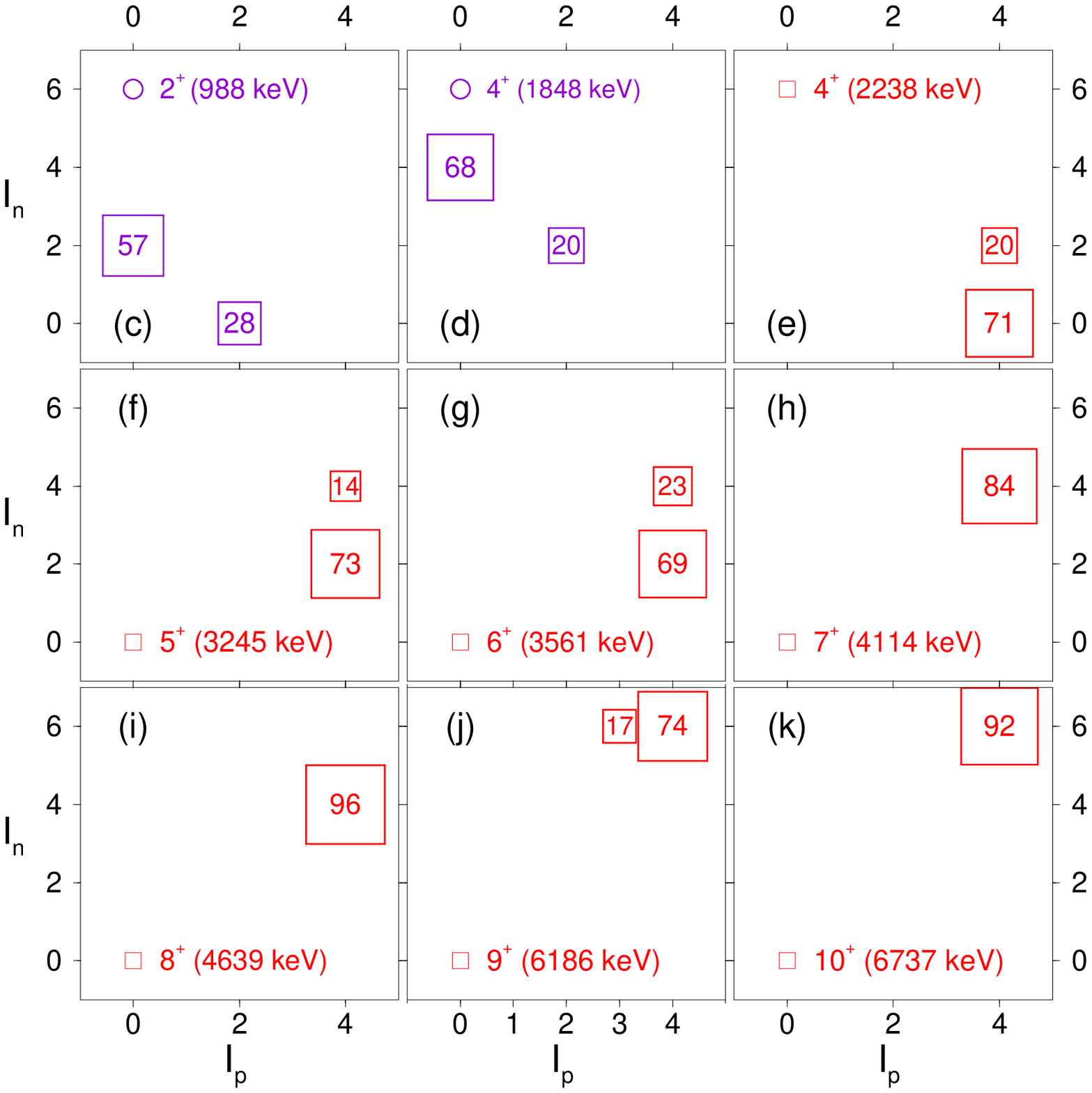}
\includegraphics[angle=-90,width=8.8cm]{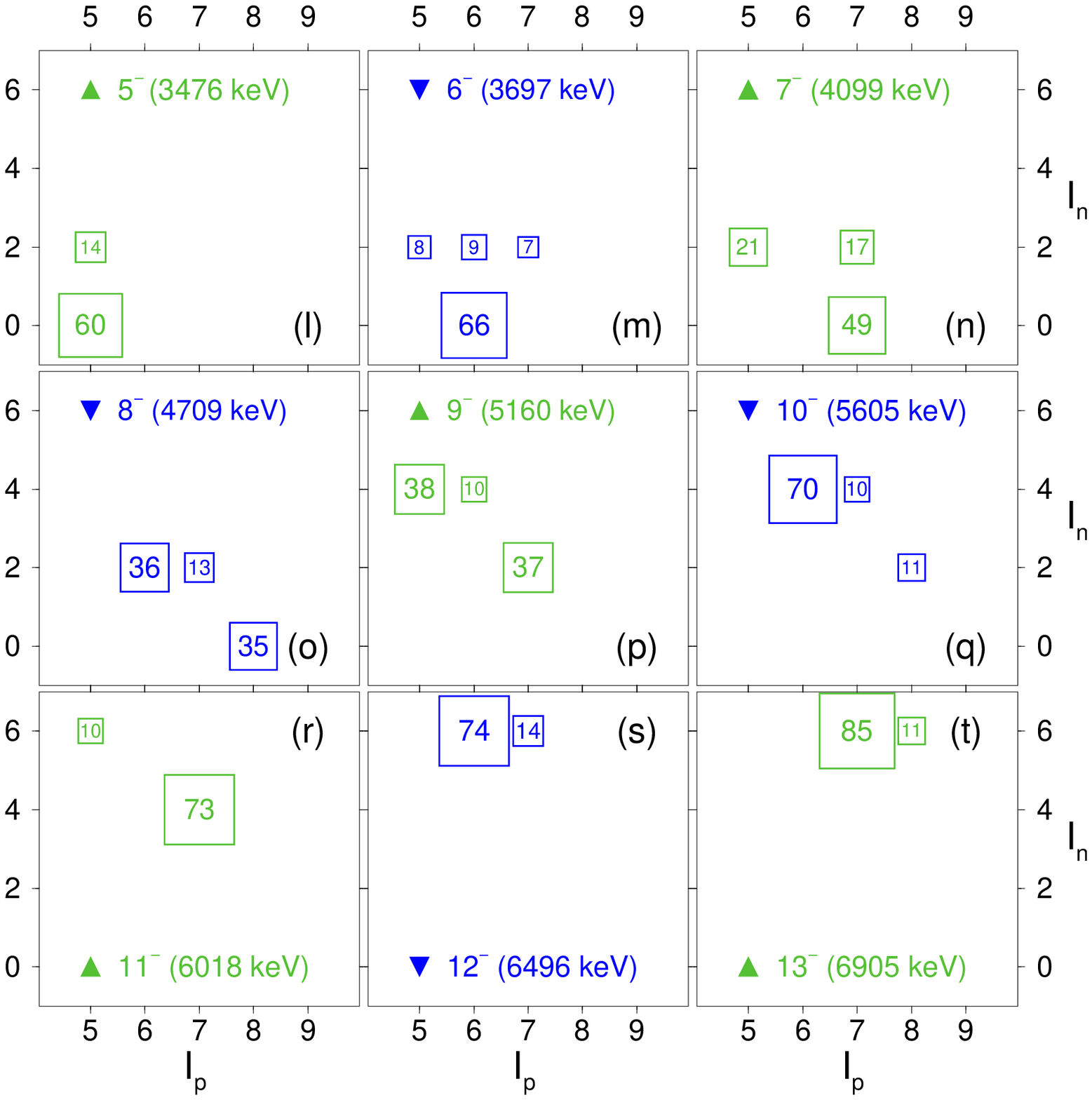}
\caption[]{(Color online) 
(a) Results of the shell-model calculations for $^{88}$Kr (see text). The states having 
the same main configuration are linked by a solid line. 
(b) Experimental states of $^{88}$Kr drawn with the same symbols as in (a).
(c--t) For each state of $^{88}$Kr drawn in (a), decomposition of its total angular 
momentum into the $I_p \otimes I_n$ components. Each percentage is written inside a 
square, drawn with an area proportional to it. Low percentages are not displayed.
(c--k) positive-parity states, (l--t) negative-parity states. }
\label{88Kr_SM_exp}
\end{figure*}

The results of the SM calculations for $^{88}$Kr are given in Fig.~\ref{88Kr_SM_exp}(a), showing only
the states close to the yrast line.
Above 2~MeV excitation energy, three sets of states are
predicted. The first one starts at $I^\pi=4^+$ and ends at $I^\pi=10^+$ [see the red empty squares in
Fig.~\ref{88Kr_SM_exp}(a)]. The two others have a negative parity, from $I^\pi=5^-$ to $I^\pi=13^-$
(see the green filled up-triangles) and from $I^\pi=6^-$ to $I^\pi=12^-$
(see the blue filled down-triangles). 

The analysis of the wave functions allows us to identify which nucleon pairs are broken. For that
purpose,  we use (i) the values of the two components, $I_p$ and $I_n$, which are coupled to give 
the total angular momentum of each state and (ii) for each $I_p$ and $I_n$ component, its decomposition in terms of proton-neutron 
configurations, i.e., the occupation numbers of the eight valence orbits which are considered in the 
present calculations. Figs.~\ref{88Kr_SM_exp}(c-t) display the decomposition of the total 
angular momentum into the $I_p \otimes I_n$ components for
each state of $^{88}$Kr drawn in Fig.~\ref{88Kr_SM_exp}(a). Each percentage 
is written inside a square, drawn with an area proportional to it. 

The major component of the 2$^+_1$ and 4$^+_1$ states comes from the breaking of the 
neutron pair, with $I_n$~=~2 (57\%) and 4 (68\%), respectively 
(see Figs.~\ref{88Kr_SM_exp}(c-d), in violet). The breaking of a proton pair (with $I_p$~=~2) 
is also involved in these two wave functions, the components being 28\% and 20\%. 

The major part of the wave functions of the set of states from $I^\pi=4^+_2$ to 
$I^\pi=10^+$ (see Figs.~\ref{88Kr_SM_exp}(e-k) in red) has $I_p$~=~4, coming from the first breaking of a proton pair, 
$(\pi f_{5/2})^1(\pi p_{3/2})^1$. In addition, the neutron
component increases from $I_n$~=~0 (for the 4$^+_2$ state) to  $I_n$~=~6 
(for the 9$^+$ and 10$^+$ states).
The 6$^+$ state of this set is the second one, the first one being predicted to
lie at 3.316~MeV [see the black star in Fig.~\ref{88Kr_SM_exp}(a)]. The major 
component (89\%) of 
this 6$^+_1$ state comes from two breakings, $I_n=4 \otimes I_p=2$.

The two sets of negative parity states have different proton configurations 
[see Figs.~\ref{88Kr_SM_exp}(l-t)]. The main components of the states with 
even-$I$ (in blue) have an even value of $I_p$, while those of the odd-$I$ 
states (in green) have an odd value 
of $I_p$.  The increase of the total angular momentum of the states belonging to the two sets comes 
from the breaking of the neutron pair, from $I_n$~=~0 to $I_n$~=~6. Then the maximum values of the angular
momentum for the two sets are $I^\pi=12^-$ and $I^\pi=13^-$, respectively. 

The experimental energies (see Fig.~\ref{schema88Kr}), drawn in Fig.~\ref{88Kr_SM_exp}(b)
using the same symbols as those of Fig.~\ref{88Kr_SM_exp}(a), show a good agreement 
with these predictions. Thus all the excited states lying above 2~MeV excitation 
energy mainly involve first the breaking of a proton pair ($I_p^\pi=4^+$ for the
positive-parity states and $I_p^\pi=5^-$--$7^-$ for the negative-parity ones), the 
increase of angular momentum inside each structure coming from the breaking of a 
neutron pair.

Nevertheless, it is important to note that the quasi-vibrational behavior of 
the first two excited states (drawn in violet) is not very well reproduced,
since the energy of the $2^+_1$ state is predicted too high (988~keV instead of 775~keV). This is
likely due to the truncation of the model space which hinders many of the correlations needed to
reproduce the low energy of this first excited state. 

Noteworthy is the fact that the 9$^-_2$ state, predicted by the calculations to lie 
at 5543~keV, has a very peculiar
wave function. It shows many different components (25\% of $2^+_n \otimes 7^-_p$,  
24\% of $4^+_n \otimes 6^-_p$, 20\% of $4^+_n \otimes 5^-_p$, 8\% of $2^+_n \otimes 8^-_p$, 
7\% of $0^+_n \otimes 9^-_p$, ...), thus resembling  a 'collective' state. It could be compared to the 
(9$^-$) experimental state at 4857~keV, i.e., 937~keV above the 7$^{(-)}$ state 
(see Fig.~\ref{schema88Kr}), even though it is predicted
too high in energy. One could similarly assume that it is due to the truncation of 
the model space. 

Moreover, a 14$^-$ state, with $I_p=8 \otimes I_n=6$ (87\%) is predicted at 7963-keV 
excitation energy\footnote{The $I_p$~=~8
component comes from the breaking of two proton pairs, i.e., the partially aligned states of the
two configurations, $(\pi f_{5/2})^2(\pi p_{3/2})^1(\pi g_{9/2})^1$ and 
$(\pi f_{5/2})^1(\pi p_{3/2})^2(\pi g_{9/2})^1$. The wave function of the 13$^-$ state at 6905~keV 
contains also such a $I_p$~=~8 component [see Fig.~\ref{88Kr_SM_exp}(t)].}. 
The highest-spin state measured in this experiment has an excitation energy very close to this 
prediction. Nevertheless its identification remains uncertain, as its distance in 
energy to the 13$^-_1$ level is predicted too high (1058~keV as compared to 479~keV).

\subsubsection{$^{89}$Rb}\label{SM89}

\begin{figure*}[!ht]
\includegraphics[angle=-90,width=14cm]{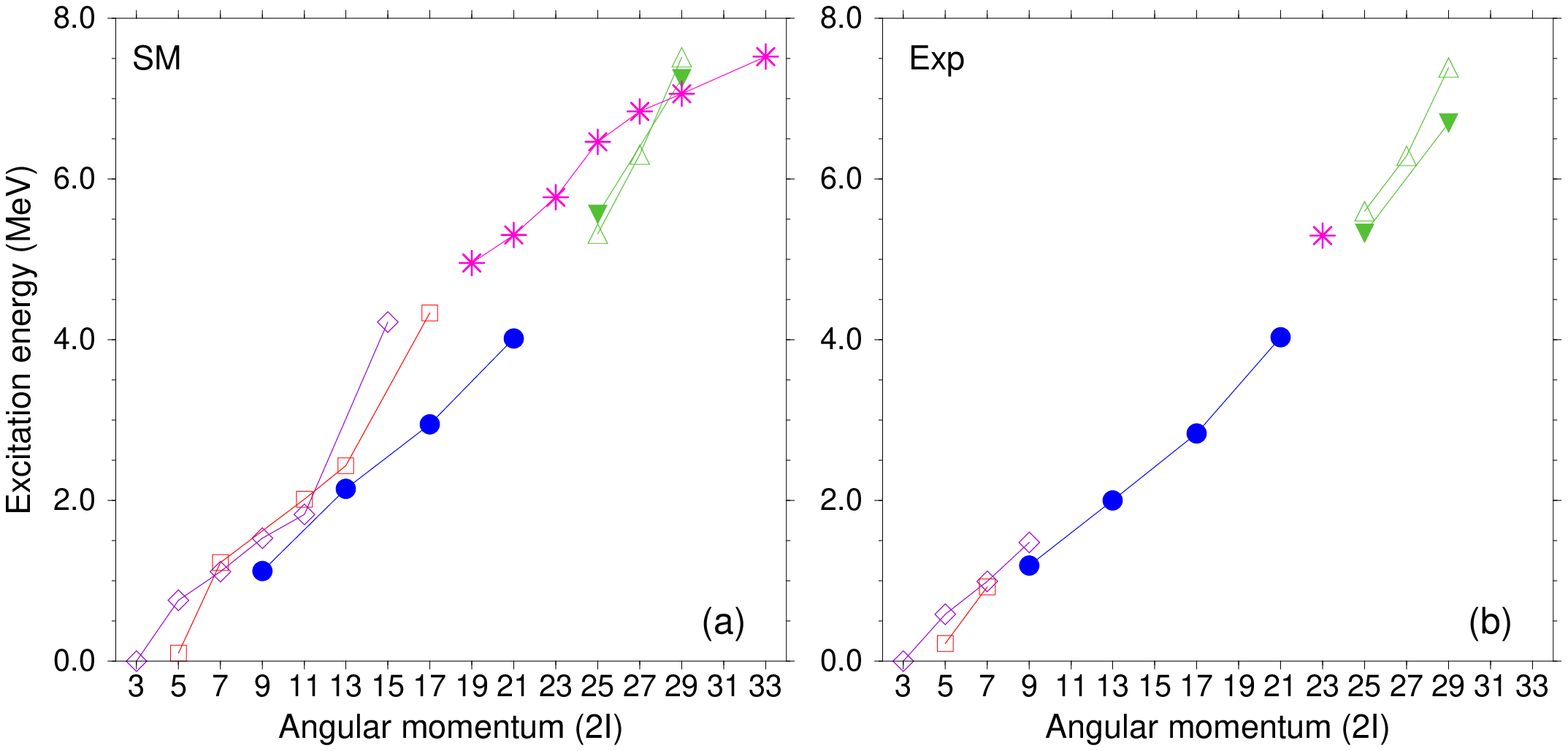}
\includegraphics[angle=-90,width=8.8cm]{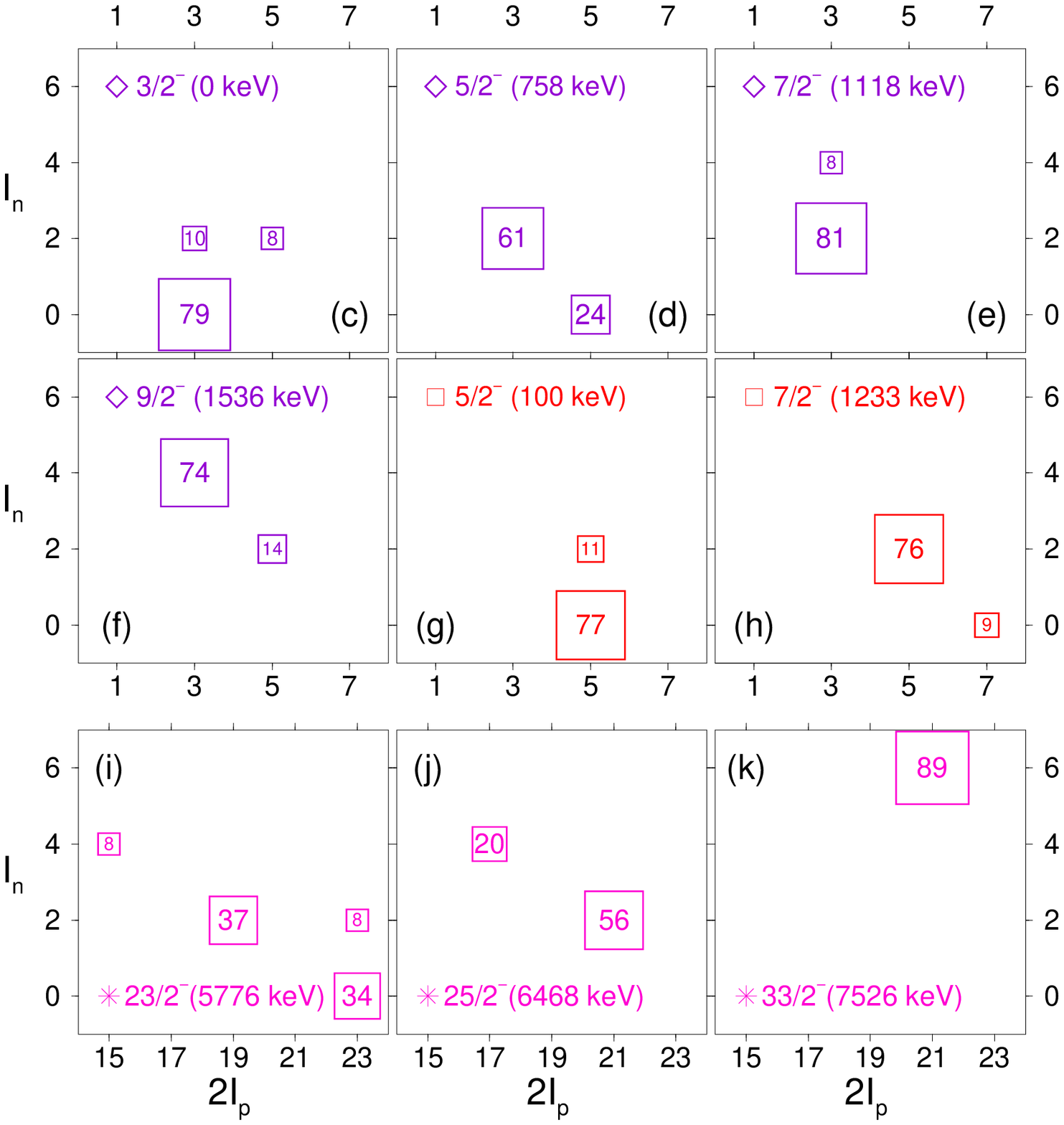}
\includegraphics[angle=-90,width=8.8cm]{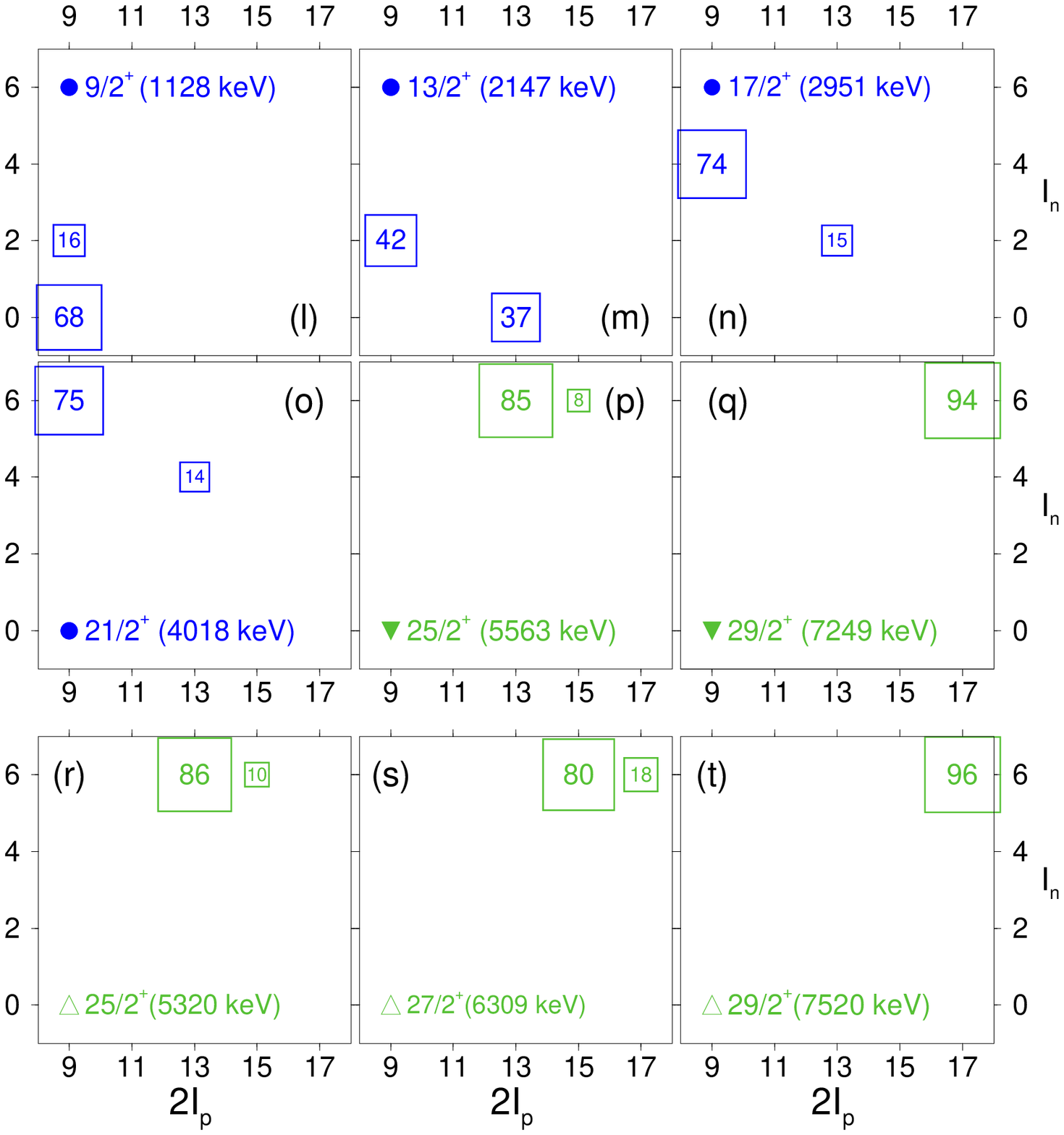}
\caption[]{(Color online) (a) Results of the shell-model calculations for $^{89}$Rb (see text). The states having 
the same main configuration are linked by a solid line. 
(b) Experimental states of $^{89}$Rb drawn with the same symbols as in (a).
(c--t) For most of the states of $^{89}$Rb drawn in (a), decomposition of their total angular 
momentum into the $I_p \otimes I_n$ components. Each percentage is written inside a 
square, drawn with an area proportional to it. Low percentages are not displayed.
(c--k) negative-parity states, (l--t) positive-parity states. 
}
\label{89Rb_SM_exp}
\end{figure*}
The results of the SM calculations for $^{89}$Rb are given in 
Fig.~\ref{89Rb_SM_exp}(a). As expected, the low-energy part of the level 
scheme displays three structures, built on the one-proton configurations, 
$(\pi p_{3/2})^1$ (see the violet diamonds), $(\pi f_{5/2})^1$ (see the red
squares), and $(\pi g_{9/2})^1$ (see the blue circles).  The maximum values 
of the angular momentum, which are obtained when breaking the neutron pair 
($I_n=6$ from the $(\nu d_{5/2})^1(\nu g_{7/2})^1$ configuration), are 
15/2$^-$, 17/2$^-$, and 21/2$^+$, respectively. 

In order to increase the angular momentum, another pair has to be broken. 
Three sets of states are then predicted at excitation energy higher than 5~MeV 
[see the magenta stars and the green triangles in Fig.~\ref{89Rb_SM_exp}(a)]. 
Since their 29/2 states are close in energy, states from these three sets 
could be identified experimentally.  

The coupling of the two components, $I_p$ and $I_n$, to obtain the total angular 
momentum of the calculated states, are drawn in Figs.~\ref{89Rb_SM_exp}(c-t), 
which allows us to sort the
states as a function of the $I_p$ value of their main component.
For instance, the band built on the 3/2$^-$ ground state (79\% with $I_p=3/2$)
comprises the 5/2$^-_2$ (61\%), 7/2$^-_1$ (81\%), and 9/2$^-_1$ (74\%) states 
(see Figs.~\ref{89Rb_SM_exp}(c-f), in violet). 
Similarly the main components of the 5/2$^-_1$ and 7/2$^-_2$ state have $I_p=5/2$ with 77\% and 76\%,
respectively (see Figs.~\ref{89Rb_SM_exp}(g-h), in red). Regarding 
the positive-parity states above the 9/2$^+$ state at 1128~keV, their main components have 
$I_p=9/2$ (see Figs.~\ref{89Rb_SM_exp}(l-o), in blue).  

The angular momentum couplings of the five positive-parity states lying above
5~MeV excitation energy are given in Figs.~\ref{89Rb_SM_exp}(p-t) (in green).
The wave functions of the two 25/2$^+$ states (predicted at 5563 and 5320~keV) have very similar 
components, the major one corresponding to $I_p=13/2 \otimes I_n=6$ (86\% and 85\%,
respectively) and the minor one to $I_p=15/2 \otimes I_n=6$ (10\% and 8\%,
respectively). This is the same for the two 29/2$^+$ states (calculated at 7249 and 7520~keV), which exhibit 
a $I_p=17/2 \otimes I_\nu=6$ 
coupling (94\% and 96\%, respectively). Nevertheless these $I_p$ values do not originate from 
the same
proton configurations, as shown in Table~\ref{decompose}, allowing us to associate the 25/2$^+_1$, 
27/2$^+_1$, and  29/2$^+_2$ states on the one hand [see the green empty up-triangles in Fig.~\ref{89Rb_SM_exp}(a)],
and the 25/2$^+_2$ and  29/2$^+_1$ states on the other hand 
[see the green filled down-triangles in Fig.~\ref{89Rb_SM_exp}(a)]. It is worth noting that
the $\pi f_{5/2}$ and  $\pi p_{3/2}$ orbits contain an odd number of protons for the main 
proton configuration of the 25/2$^+_1$, 27/2$^+_1$, and  29/2$^+_2$ states, while they
contain an even number of protons for the other set of states. 
\begin{table*}[!ht]
\caption{Proton configurations of the positive-parity high-energy states of $^{89}$Rb. Percentages greater 
than 5\% are given and the greatest value of each state is underlined.
}
\label{decompose}
\begin{ruledtabular}
\begin{tabular}{lccccc}
Proton configuration 					&25/2$^+_1$ &25/2$^+_2$ &27/2$^+_1$ &29/2$^+_1$ &29/2$^+_2$\\
		     						&5320~keV &5563~keV &6309~keV &7249~keV &7520~keV \\
\hline
&&&&&\\
$(\pi f_{5/2})^5(\pi p_{3/2})^3(\pi p_{1/2})^0(\pi g_{9/2})^1$& \underline{32\%} &14\%&\underline{53\%}&11\%&\underline{41\%}\\
$(\pi f_{5/2})^4(\pi p_{3/2})^4(\pi p_{1/2})^0(\pi g_{9/2})^1$&	19\%&\underline{39\%}& &\underline{53\%}&13\%	\\
$(\pi f_{5/2})^5(\pi p_{3/2})^2(\pi p_{1/2})^1(\pi g_{9/2})^1$&	8\%	&9\%	&20\%	& 	&21\%	\\
$(\pi f_{5/2})^4(\pi p_{3/2})^3(\pi p_{1/2})^1(\pi g_{9/2})^1$&	12\%	&7\%	&10\%	&17\%	&10\%	\\
$(\pi f_{5/2})^3(\pi p_{3/2})^4(\pi p_{1/2})^1(\pi g_{9/2})^1$&		&10\%	&	&	&	\\
\end{tabular}
\end{ruledtabular}
\end{table*}

Regarding the third set of states predicted above 5~MeV excitation energy 
[see the magenta stars in Fig.~\ref{89Rb_SM_exp}(a)], their wave functions 
are characterized by the occupation of the $\pi g_{9/2}$ orbit by two 
protons, the pair being broken so as to give their high values of $I_p$ 
(see Figs.~\ref{89Rb_SM_exp}(h-k), in magenta). This set 
of states extends from $I^\pi=33/2^-$ down to $I^\pi=19/2^-$, the bottom 
of the structure being located well above the yrast line. Then, one can 
assume that only the high-energy part would be populated in experiments 
such as the present one. For instance, the 23/2$^-$ state decays more 
likely to the 21/2$^+_1$ state than to the 21/2$^-$ state, because of
the largest $\gamma$-ray energy.

The experimental energies (see Fig.~\ref{schema89Rb}) drawn in 
Fig.~\ref{89Rb_SM_exp}(b) using the same symbols as those of 
Fig.~\ref{89Rb_SM_exp}(a), show a good agreement with these predictions. 
In the bottom of the level scheme, the odd proton lies in one of the 
odd-parity orbit. For $I \ge 9/2$, the positive-parity states involving 
the occupation of the $\pi g_{9/2}$ orbit form the yrast line. This 
structure ends at $I^\pi=21/2^+$ where the two neutrons bring the maximum 
angular momentum, $6\hbar$. The greater values of the angular momentum 
are then obtained by breaking a pair of proton. Three sets of states are
predicted with different proton configurations, in agreement with the 
observation of three transitions populating the 21/2$^+$ state at 
4032~keV (see Fig.~\ref{schema89Rb}).  

\subsubsection{Conclusion}
The set of the "glekpn" effective interactions describes well the 
high-spin states of the two $N=52$ isotones studied in the present work. 
The only weakness seems to be the energy of the first quadrupole excitation 
which is predicted slightly too high. The $E(2^+_1 \rightarrow 0^+_1)$ 
value is 988~keV instead of 775~keV in $^{88}$Kr. A similar deviation is
observed in $^{89}$Rb when the odd proton lies in the $\pi g_{9/2}$ orbit, 
i.e., $^{89}$Rb~=~$^{88}$Kr~$\otimes (\pi g_{9/2})^1$: 
$E(13/2^+_1 \rightarrow 9/2^+_1)=$~1019~keV instead of 809~keV. A good
description of such states would need the excitations of some nucleons 
across the $Z=28$ and $N=50$ shell gaps, which were not considered in 
the present work because of the too large dimensions of the valence 
space for $Z \sim 36$ and $N=52$.

\subsection{Collectivity of the $N=52$ isotones}
As said above, the appearance of several groups of coincident transitions 
having similar energies of 800--900 keV can be considered as a sign of 
the excitation of an harmonic vibrator. This means that many $N=52$ 
isotones exhibit this collective behavior in the low-energy parts of 
their level schemes. It is the case of the even-$Z$ isotones, as well 
as the $\pi g_{9/2}$ structure of $^{89}$Rb discussed in the previous 
section. The latter can be considered as the weak coupling of a
$g_{9/2}$ proton to the states of an harmonic vibrator, which is confirmed 
by the SM calculations which predict the other members of the expected 
multiplets at excitation energies close to the one-phonon and two-phonon 
energies [being not yrast, these states are not drawn in 
Fig.~\ref{89Rb_SM_exp}(a)]. For instance the $11/2_1^+$ state is predicted
at 2129~keV, i.e., very close to the $13/2_1^+$ state, both from the 
weak coupling of the $\pi g_{9/2}$ to the one-phonon excitation. Similarly, 
the $15/2_1^+$ state is calculated at 3064~keV, that is only 113~keV above 
the $17/2_1^+$ state, both from the weak coupling of the $\pi g_{9/2}$ to 
the two-phonon excitation. 

The low-lying negative-parity states of $^{89}$Rb cannot be interpreted 
within the same approach, as they do not exhibit $\Delta I=2$ sequences, 
but $\Delta I=1$ ones (with $\Delta I=2$ cross-over transitions). On the 
other hand, these states can be discussed in terms of a 
"rotor + quasiparticle" approach, forming two rotational structures, 
built on the 3/2$^-$[301] and 5/2$^-$[303] deformed  states, respectively. 
For a deformation parameter, $\epsilon=0.15$, the $Z=37$ Fermi level is 
located on the 3/2$^-$[301] state, the 5/2$^-$[303] one lying just below 
it (see the Figure 5 of Ref.~\cite{po11}). 

It is interesting to discuss the case of the lighter $N=52$ isotone, 
$^{87}_{35}$Br. In order to explain the feeding of some medium-spin states 
of $^{87}$Kr from the $\beta$-decay of the $^{87}_{35}$Br ground 
state~\cite{po06}, its spin value has been recently reevaluated 
to be 5/2$^-$ (instead of the previous value, 3/2$^-$, chosen from 
systematics, particularly because of its neighboring semi-magic isotope, 
$^{85}$Br). Following the simple rule for the filling of the 
deformed proton orbits, the $Z=35$ Fermi level is then on the 5/2$^-$[303] 
state, which gives $I^\pi=5/2^-$ for the ground state, in perfect agreement 
with the new value. 

\begin{figure}[!h]
\includegraphics[width=8.5cm]{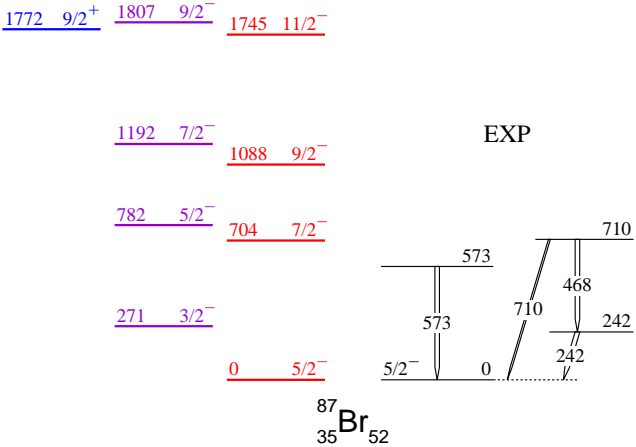}
\caption[]{(Color online) Levels of $^{87}$Br predicted by the SM 
calculations, the color code is the same as the one of Fig.~\ref{89Rb_SM_exp}(a). 
The experimental states identified from the $\beta$-decay of 
$^{87}$Se~\cite{ze80} are shown in the right part (the 5/2$^-$ value of the 
ground-state spin is from Ref.~\cite{po06}). 
}
\label{87Br}
\end{figure}

The results of the SM calculations of $^{87}$Br corroborate this spin value 
and predict two $\Delta I=1$ structures built on the ground state 
and the 3/2$^-$ first excited state (see the left part of Fig.~\ref{87Br}). 

$^{87}$Br being in the heavy-$A$ tail of the Br fragment distribution of the
fusion-fission reaction used in the present work, its population is 
unfortunately too low to identify its yrast states using our data set. 
On the other hand, a few excited states were obtained from the 
$\beta$-decay of $^{87}$Se studied many years ago~\cite{ze80}. The first 
three ones, drawn in the right part of Fig.~\ref{87Br}, compare well with 
those predicted by the SM calculations. The 242-keV state is likely
the 3/2$^-$ band head, while the 573- and 710-keV levels would be the 
first members of the two collective structures.

\section{Summary}
The two neutron-rich nuclei, $^{88}$Kr and $^{89}$Rb, have been produced as 
fission fragments in the fusion reaction 
$^{18}$O+$^{208}$Pb at 85-MeV bombarding energy, their $\gamma$ rays being
detected using the Euroball array. Their high-spin level schemes were
built up to $\sim$~8 MeV excitation energy by analyzing triple 
$\gamma$-ray coincidence data. Using $\gamma$-$\gamma$ angular correlation
results, spin and parity values were assigned to most of states observed in  
the two isotones below 4-MeV excitation energy. The behaviors of the yrast 
structures identified in this work have been discussed, first, in comparison 
with the general features known in the mass region. Then results of SM 
calculations using the "glekpn" effective interactions in the 
$\pi f_5pg_9 \otimes \nu g_7ds$ valence space have been successfully 
compared to the experimental results. They describe well the evolution 
from collective to single-particle behaviors of these two isotones.
Thanks to the components of the wave functions, particularly the two values, 
$I_p$ and $I_n$, coupled to give the total angular momentum of each state, 
the increase of the angular momentum along the yrast lines could be 
precisely analyzed by identifying what nucleon pairs are successively 
contributing. Lastly, the behavior of the yrast states of the lighter 
isotone, $^{87}$Br, has been discussed using the SM calculations, which 
predict that its ground state has $I^\pi=5/2^-$, in agreement with a recent 
experimental result.
   
\begin{acknowledgments}
The Euroball project was a collaboration among France, the 
United Kingdom, Germany, Italy, Denmark and Sweden. 
The experiment was supported in part by the EU under contract 
HPRI-CT-1999-00078 (EUROVIV). 
We thank many colleagues for their
active participation in the experiments, Drs. A.~Bogachev, A.~Buta,  
F.~Khalfalla, I.~ Piqueras, and R. Lucas. 
We thank the crews of the Vivitron. 
We are very indebted to M.-A. Saettle
for preparing the Pb target, P. Bednarczyk, J. Devin, J.-M. Gallone, 
P. M\'edina, and D. Vintache for their help during the experiment. 
\end{acknowledgments}

\end{document}